\documentstyle[11pt,aaspp4,epsfig]{article}

\def\beq{\begin{equation}}
\def\enq{\end{equation}}
\def\beqn{\begin{eqnarray}}
\def\enqn{\end{eqnarray}}

\begin{document}

\title{Spectral properties, generation order parameters and luminosities for spin-powered X-ray pulsars}
\author{Wei Wang and Yongheng Zhao}
\begin{center}
National Astronomical Observatories, Chinese Academy of Sciences, Beijing  100012, China \\
e-mail: wwang@lamost.bao.ac.cn, yzhao@lamost.bao.ac.cn
\end{center}

\begin{abstract}
We show the spectral properties of 15 spin-powered X-ray pulsars, and the correlation between the average power-law photon index and spin-down rate. Generation order parameters (GOPs) based on polar-cap models are introduced to characterize the X-ray pulsars. We calculate three definitions of generation order parameters due to the different effects of magnetic and electric fields on photon absorption during cascade processes, and study the relations between the GOPs and spectral properties of X-ray pulsars. There exists a possible correlation between the photon index and GOP in our pulsar sample. Furthermore, we present a method due to the concept of GOPs to estimate the non-thermal X-ray luminosity for spin-powered pulsars. Then X-ray luminosity is calculated in the context of our polar-cap accelerator model which is well consistent with the most observed X-ray pulsar data. The ratio between X-ray luminosity estimated by our method and the pulsar's spin-down power is well consistent with the $L_{\rm X}\sim 10^{-3}L_{\rm sd}$ feature. 

\end{abstract}


\section{Introduction}
X-ray study of the spin-powered pulsars has increased substantially recently. In a long time, Crab pulsar was the just one radio pulsar detected at X-ray energies (Toor \& Seward 1977). The first big step was taken by Einstein Observatory which discovered X-ray pulsations from two Crab-like pulsars PSR 0540-69 and 1509-58 (Seward \& Wang 1988). Recently, X-ray observations like ROSAT, ASCA, RXTE and BeppoSAX have achieved an important progress in neutron star and pulsar astronomy (Becker \& Tr\"umper 1997, Strickman et al. 1999), including more pulsations detected, fluxes comfined and spectral properties (thermal and non-thermal components). At present, the advanced X-ray missions like Chandra and XMM-Newton provide us more detections specially of faint X-ray emitters and high resolution observations in the field of pulsar astronomy.

The X-ray radiation of spin-powered pulsars attributes to thermal and non-thermal emission processes including: non-thermal emission from relativistic particles accelerated in the magnetosphere (Michel 1991); photospheric emission from the hot surface of cooling neutron star (Kundt \& Schaaf 1993); thermal emission from hot polar caps (Gil \& Krawczyk 1996); extended radiation from synchrotron nebulae (Michel 1991); inverse Compton scattering of softer photons (Zhang \& Harding 2000) and so on. In the present paper, we concentrate on the emission from relativistic particles accelerated in the pulsar magnetosphere, which is characterized by a power-law spectrum, described by a power-law photon index. Here, we find an interesting correlation between the photon index and spin-down rate of the pulsars.

There exist many magnetosphere emission models to explain the high energy radiation ($\gamma$ and X-rays) from spin-powered pulsars. A general scenario is proposed by outer gap models (Cheng, Ho \& Ruderman 1986a, 1986b, Wang et al. 1998, Zhang \& Cheng 2000) to produce the $\gamma$ and X-ray photons. In this model, an outer-gap accelerator send $e^+/e^-$ pairs flowing inward and outward along open magnetic field lines. These particles radiate $\gamma$-ray by the curvature mechanism. When inward flowing particles approach the star surface, curvature emitting $\gamma$-ray photons greater than 100 MeV convert into secondary $e^+/e^-$ in the inner magnetosphere where $B\sin\phi > 2\times 10^{10}$G, $\phi$ is the angle between the photon and the magnetic field. The secondary $e^+/e^-$ will radiate through synchrotron mechanism. We also introduce a concept of generation order parameter (GOP) which originally described the cascade processes of $\gamma$-ray pulsars (Zhao et al. 1989, Lu, Wei \& Song 1994, Wei, Song \& Lu 1997). GOPs are studied in the scenario of polar cap models (Ruderman \& Sutherland 1975). And it is found that the GOPs are correlated with $\gamma$-ray spectral index of $\gamma$-ray pulsars (Lu et al. 1994, Wei et al. 1997). We use GOPs to study X-ray pulsars, and think that the different photon indexes could be characterized by the GOPs if their radiation is really described by the cascade processes in the same emission mechanism. Due to the scenario of GOPs, we also provide another method to estimate the X-ray luminosity of spin-powered pulsars (Zhao et al. 1994). We find that our GOP method based on poler-cap models could well predict the X-ray luminosity of Crab-like pulsars and millisecond pulsars.

This paper will be organized as follows. In Section 2, we will show our sample of spin-powered X-ray pulsars and their spectral properties. Generation order parameters are introduced in Section 3, and the relations between different GOPs and X-ray photon indexes are displayed. In Section 4, we calcalate X-ray luminosity of these spin-powered pulsars in our polar-cap model and compare them to the observational values. Summary and discussions are outlined in the final section.

\section{X-ray pulsar sample and spectral properties}
We are interested in the non-thermal components and spectral properties in the X-ray radiation of spin-powered pulsars. Then we just select 15 X-ray pulsars with the spectral measurements by high resolution detectors (see Table 1). The spectral feature is a power-law of the average photon index $\Gamma$ which varies from 1.5 (Crab) to 3.0 (PSR J0030+0451). In our sample, there are 7 gamma-ray pulsars (Crab, Vela, Geminga, PSR 1706-44, 1951+32, 2229+6114, 1055-52), where PSR 2227+6122 is the new discovered $\gamma$-ray and X-ray pulsar identified by ASCA and EGRET (Halpern et al. 2001a, 2001b), and three millisecond pulsars (PSR 1821-24, 0437-47, 0030+0451). In our X-ray pulsar sample, we have taken the latest observational results which help our following analysis. The recent observational results of Vela by RXTE are shown by Harding et al. (2002), the pulsed X-ray emission of PSR 1706-44 is detected by Chandra (Gotthelf et al. 2002). Chandra also reported the spectral and time observations of the nearest millisecond pulsar J0437-4715 (Zavlin et al. 2002). The observational data of PSR 1055-52 and 0030+0451 is provided by recent observations of XMM-Newton (Becker \& Aschenbach 2002). The other pulsars' observations are reported by both ROSAT and XMM-Newton (Becker \& Tr\"umper 1997, Becker \& Aschenbach 2002). We'd like to emphasize that the observed X-ray luminosities of 15 pulsars in Table 1 include only the pulsed luminosities possibly coming from non-thermal components.

The power-law emission components of spin-powered X-ray pulsars come from relativistic particles accelerated in pulsar magnetosphere, whose energy originates from the pulsar's rotational energy loss. Then the spectral properties may be closely related to the pulsar's spin characteristics. Figure 1 shows the average power-law photon index for each X-ray pulsar as a function of the spin-down rate $|\dot\Omega|=2\pi \dot P/P^2$ which is proportional to the spin-down torque. We find an approximate correlation between $\Gamma$ and $|\dot\Omega|$ with a large dispersion. As shown in Figure 1, the spectra of spin-down X-ray pulsars become harder with the spin-down rate increasing. The similar relation has also been found in X-ray properties of anomalous X-ray pulsars and softer gamma-ray repeaters (Marsden \& White 2001, Lu, Wang \& Zhao 2003). It implies that the same mechanism might induce this common correlation.

For a further analysis, we let the photon index as a function of $P$ and $\dot P$, $\Gamma(P,\dot P)$. We fit the relation in terms of a linear function $\Gamma=\alpha \log P+ \beta \log\dot{P}+\gamma$, $\Gamma(P,\dot P)$ of 15 X-ray pulsars which is given as
\beq
\Gamma(P,\dot P)\approx -0.55(\pm 0.1){\rm log}P+0.21(\pm 0.05){\rm log}\dot{P}_{15}+0.42(\pm 0.3).
\enq
The right-hand side of the above equation has a similar term to the definition of generation order parameters in the following section. So there exits a possible relation between photon indexes and generation order parameters for spin-powered X-ray pulsars, which is discussed detailedly in Section 3.

\section{Generation order parameters for X-ray pulsars}J
The concept of generation is originally provided to describe the cascade processes in $\gamma$-ray pulsars (Zhao et al. 1989, Lu \& Shi 1990). Based on Ruderman-Sutherland model (Ruderman \& Sutherland 1975), passing through the polar gap, $e^+/e^-$ are accelerated to a high energy with typical Lorentz factor $\gamma_1=6.0\times 10^7 P^{1/14}\dot{P}_{15}^{-1/14}$, where $P$ is the period of pulsar in units of second, $\dot{P}_{15}$ the derivative in units of $10^{-15}$s\ s$^{-1}$. These first generation particles will move along the curved magnetic field lines and emit high energy curvature radiation (the first generation photons) with photon energy typically concentrated at
\beq
E_c={3\over 2}{\hbar c\over R_c}\gamma_1^3\approx 3.2\times 10^{10}P^{-2/7}\dot{P}_{15}^{-3/14}{\rm eV},
\enq
where $R_c\approx 1.8\times 10^7 P^{1/2}$cm is the curvature radius of field line here. According to Hardee (1977), the $\gamma$-ray photon energy higher than the critical energy
\beq
E_a=9.5\times 10^9 P^{1/2}\dot{P}_{15}^{-1/2}{\rm eV}
\enq
will be absorbed and transformed into $e^+/e^-$. The Lorentz factor $\gamma_2$ of the second generation $e^+/e^-$ will be $E_1/{2m_e c^2}$. These $e^+/e^-$ can emit the second generation photons through synchrotron radiation with a characteristic energy $E_2$. If $E_2 > E_a$, further $e^+/e^-$ can be produced, then cascade processes occur.

Concerning this idea, Lu et al. (1994) introduced the generation order parameter (GOP) to characterize a pulsar. They considered the conversion of high energy photons into $e^+/e^-$ pairs through electric fields, and defined the GOP as
\beq
\zeta_1=1+{1-(11/7){\rm log}P+(4/7){\rm log}\dot{P}_{15}\over 3.56-{\rm log}P-{\rm log}\dot{P}_{15}}.
\enq
The GOPs are proved to be correlated with the $\gamma$-ray photon index and the mean photon energy: softer $\gamma$-ray photons with larger GOPs. This relation is consistant with the scheme of cascade processes.

Wei et al. (1997) took a different approximation from Lu's scheme. Here, high energy photons are absorbed through the effect of both magnetic and electric fields. The second GOP is defined as follows,
\beq
\zeta_2=1+{0.8-(2/7){\rm log}P+(2/7){\rm log}\dot{P}_{15}\over 1.3}.
\enq

The concept of generation was initially considered in the scheme that the $\gamma$-ray photons is absorbed and conversed into $e^+/e^-$ through only magnetic fields (Zhao et al. 1989). Then we should define a new generation order parameter in this scheme. Since $E_c=E_a 20^n$, we obtained
\beq
\zeta_3=1+n=1+{0.6-(11/14){\rm log}P+(2/7){\rm log}\dot{P}_{15}\over 1.3}.
\enq

GOPs are used to describe cascade processes and characterize the spectral properties of pulsars. Then we plot the diagrams between the GOPs and photon index of 15 X-ray pulsars, and try finding the relation of $\zeta-\Gamma$ which is implied in Eq. 1. The photon index as the function of the GOPs $\zeta_{1,2,3}$ is displayed in Figs. 2, 3 and 4, respectively. The figures all show a possible correlation between GOPs and photon indexes. In X-ray bands, the photon index $\Gamma$ becomes smaller with increasing the generation order parameter $\zeta$. This relation is different from the one derived in $\gamma$-ray bands (Lu et al. 1994, Wei et al. 1997). Three GOPs are all displayed here for a comparasion. One can find $\zeta_3-\Gamma$ shows the better correlation than the others. According to the different definitions of three GOPs, we draw a conclusion that in X-ray pulsars, magnetic field may have the most significant effect on the high energy photon absorption during cascade processes, which is also consistent with the conclusion of a theoretical work by Zheng, Zhang \& Qiao (1998), in which they have shown that although Daugherty \& Lerche's unified formula (1975) for gamma-ray absorption involving both E and B is correct, the E field is not important at all for pair production in the pulsar magnetosphere. The main point is that the photon direction in the laboratory frame used by Daugherty \& Lerche (1975) is wrong. Calculations of pure E case (Lu et al. 1994) and E+B case (Wei et al. 1997) are based on the wrong application of Daugherty \& Lerche's formula, and then only $\zeta_3$ (pure B case by Zhao et al.1989) is the appropriate parameter. Therefore, this may be the reason why there is a tighter relation between $\zeta_3$ and the photon index in Fig. 4.

\section{X-ray luminosity of spin-powered pulsars}
The theoretical models for X-ray radiation from spin-powered pulsars are still the open problems. In this paper, we concentrate on the non-thermal radiation from X-ray pulsars considering the emission from relativistic particles accelerated in pulsar magnetosphere. Generally, magnetospheric accelerators fall into two main classes: outer gap and polar cap models. In this section, we will emphasize our polar cap model according to the concept of generation order parameters.

\subsection{Polar cap models and generation order parameters}
Generation order parameters based on polar cap models (Ruderman \& Sutherland 1975) have been used to estimate $\gamma$-ray luminosity of $\gamma$-ray pulsars (Zhao et al. 1989, Lu et al. 1994, Wei et al. 1997). Here, we propose that the concept of GOPs can also be used to calculate X-ray luminosity of spin-powered pulsars. 

In polar cap region of a pulsar, the maximum surface electric potential is (Goldreich \& Julian 1969)
\beq
\Delta V_{\rm max}={\Omega^2 BR^3\over 2c^2}.
\enq
Then the maximum flowing luminosity by the primary particles (the first generation) is given as (Ruderman \& Sutherland 1975)
\beq
L_{\rm max}=\dot{N}_0 e\Delta V_{\rm max}\approx 6.1\times 10^{31}P^{-3}\dot{P}_{15}R_{6}^6 {\rm erg\ s^{-1}},
\enq
where $\dot{N}_0\sim \Omega^2 B_p R^3/ec$ is the maximum current through the accelerator, $B_p\sim 2\times 10^{12}P^{1/2}\dot{P}_{15}^{1/2}$G is the surface dipole field, $R$ is neutron star radius about $10^6$cm. Though Harding \& Muslimov (2002) have argued that the above equation is not exact for young pulsars whose accelerator are limited by pairs which screen the parallel electric field at some attitude, we still think the formula is valuable for all pulsars in our calculations.

As discussed in Section 3, only the photons with the energy lower than $E_a$ can escape from pulsar magnetosphere. Assuming a power-law energy spectrum after the cascade processes from the pulsar (Zhao et al. 1994): $f(E)=AE^s$, where A is a constant for one pulsar. Here, we should note that from X-ray to $\gamma$-ray range, the observations have found many pulsars have the spectral breaks, but the average spectral index we defined here is still meaningful. The average spectral index would be correlated with generation order parameters in theories, and in addition, the index can help us to estimate the non-thermal X-ray luminosity of X-ray pulsars. So the total luminosity from a pulsar is
\beq
L_{\rm max}=\int_0^{E_a}f(E)dE={A\over 1+s}E_a^{1+s},
\enq
here we have assumed $1+s>0$. In the same way, we obtain the X-ray luminosity as
\beq
L_{\rm X,GOP}=\int_{E_1}^{E_2}f(E)dE={A\over 1+s}(E_2^{1+s}-E_1^{1+s}),
\enq
where $E_1,E_2$ are lower and upper limits of X-ray bands (here, we take $E_1 = 0.1$ keV, and $E_2 = 2.4$ keV in our calculations, which is convenient to compare the prediction to observed value.). With Eqs. 9 and 10, we find the relation between X-ray luminosity and total luminosity for a spin-powered pulsar:
\beq
L_{\rm X,GOP}=L_{\rm max}{E_2^{1+s}-E_1^{1+s}\over E_a^{1+s}}.
\enq

From Eq. 11, we could calculate X-ray luminosity for a pulsar. Firstly, we should check the average high energy spectrum $s$ using the observed data of seven $\gamma$-ray pulsars in our sample. $\gamma$-ray luminosity in our scheme is given as
\beq
L_\gamma=\int_{E_3}^{E_a}f(E)dE={A\over 1+s}(E_a^{1+s}-E_3^{1+s}),
\enq
$E_3$ is lower energy limit for $\gamma$-ray observation ($E_3=100$MeV for EGRET, and we take $E_3=10$keV for Crab pulsar). Then using Eqs. 10 and 12, we have
\beq
{L_\gamma\over L_{\rm X}}={E_a^{1+s}-E_3^{1+s}\over E_2^{1+s}-E_1^{1+s}}.
\enq
We take the observed $\gamma$ and X-ray luminosities of 7 $\gamma$-ray pulsars into Eq. 13 to derive an average energy spectrum $s$ from $\gamma$-ray to X-ray bands, where $\gamma$-ray luminosity of Crab is taken from Buccheri (1981) and others from McLaughlin \& Cordes (2000). Finally, we find $s=-0.6\pm 0.4$ (refer to Fig. 5). So we let $s=-0.6$ for all 15 pulsars and calculate X-ray luminosities of these pulsars.(Table 1). The observed and calculated values are also compared, the results are displayed in Fig. 6. The solid line represents the diagonal line $L_{\rm X,obs}=L_{\rm X,GOP}$, the dashed line is the best fitting line $L_{\rm X,obs}=0.02L_{\rm X,GOP}^{1.02}$. Our theoretical prediction is consistent with the observational value, but Vela pulsar can not be well explained in our scheme.

In theory, we expect the steeper average energy spectrum with larger generation order parameters. Then we derive each average high energy spectrum for 7 $\gamma$-ray pulsars, and plot them in Fig. 5 as the fuction of the generation order parameter $\zeta_3$. The dashed line denotes $s=-0.6$. The data points show that the energy spectrum is nearly independent of $\zeta_3$, and $s$ value is really around the dashed line, which implies our consistancy of using $s=-0.6$ in all pulsars. The $s$ value of Crab is smallest with the largest generation order parameter. If comparing Crab with other pulsars, a possible weak correlation between $s$ and $\zeta_3$ exists, softer energy spectra with increasing generation order parameters which is consistant with the interpretation of cascade processes.

Our GOP model based on the polar-cap scenario suggests that the primary energy source of X-ray emission is the pulsar's rotational energy. Fig. 8 shows the X-ray luminosity of 15 pulsars in our sample estimated by our polar-cap model ($s=-0.6$) as a function of their rotational energy loss $L_{\rm sd}$. Furthermore, we also plot the observed X-ray luminosity of pulsars in 0.1-2.4 keV range as the function of their spin-down energy in Fig. 7. The 14 pulsars' rotational energy is taken from Becker \& Tr\"umper (1997) and Becker \& Aschenbach (2002), and the observed parameters of PSR 2227+6122 is taken from Halpern et al. (2001). In Figs. 7 and 8, each solid line represents $L_{\rm X}=10^{-3}L_{\rm sd}$, the correlation derived by the observed X-ray luminosity and the spin-down power in ROSAT energy band (Becker \& Tr\"umper 1997). For a comparison, we also plot the dashed line $L_{\rm X}=10^{-21}L_{\rm sd}^{3/2}$ derived by ASCA in the figure. We find that the correlation between our theoretical X-ray luminosity and spin-down power is well consistent with the ROSAT relation. The strong correlation between the predicted luminosity and spin-down power comfirms the conclusion of the promary X-ray emission energy of a pulsar from the rotational energy. In addition, the small dispersion of the data points around the solid line shows that our polar-cap scheme can well explain the $L_{\rm X}\sim 10^{-3}L_{\rm sd}$ feature.

\subsection{Outer-gap or polar-cap dominated?}
In this section, we have calculated X-ray luminosities of spin-powered pulsars in the context of polar cap model and compared them with observations. Our main results are displayed in Figures 5 , 6 and 8. In Fig. 6, we find the polar-cap scenario could explain luminosities of most X-ray pulsars in our sample except Vela.

From the detailed comparison in Fig. 6, we find that the predicted luminosities of most pulsars are consistent with the observed value. The special one is Vela pulsar which deviate much from the diagonal line. We further consider the outer-gap model (Cheng, Ho \& Ruderman 1986a, 1986b, Wang et al. 1998) and their predicted X-ray lumonosity, the Vela-type pulsars (including Vela pulsar, PSR 1706-44 and 1951+32) can be well explained in the outer-gap accelerator sheme. Then we conclude that the Vela-type pulsars should be outer-gap dominated. The model of Cheng, Ho \& Cheng (1986a, 1986b) could not well explain the X-ray luminosity of the Crab-like pulsars and millisecond pulsars. What's more, there are the correlations between the photon index and generation order parameters in these pulsars, and the GOPs are defined here in the context of polar-cap accelerators. Thus, we think that the Crab-like pulsars and millisecond pulsars are polar-cap dominated, which is consistent with the result of Zavlin \& Pavlov (1998). At last, we emphasize that we have not concluded that the outer-gap models can not explain the other pulsars well. Other outer-gap models ( revising the acceleration size of outer-gap region, e.g. Zhang \& Cheng 2000, Hirotani \& Shibata 2001, Hirotani, Harding \& Shibata 2003) could do better at matching the observed X-ray luminosities of pulsars.

\section{Summary and discussions}
We have presented X-ray spectral properties of 15 spin-powered pulsars obsereved by ROSAT, ASCA, RXTE, Chandra and XMM-Newton. We are interested in the non-thermal emission components, and study the relation between the average power-law photon index and spin-down rate. Generation order parameters based on polar-cap models are introduced to describe the X-ray pulsars. And the relations of three GOPs from different definitions and the photon index are also displayed. Finally, according to the concept of generation order parameters in polar-cap scenario, X-ray luminosities of spin-powered pulsars are calculated and compared with the observed data. Our main results in this paper are summarized as follows:

1. We have found a correlation between the average power-law photon index and spin-down rate in 15 X-ray pulsars: harder X-ray spectrum with increasing spin-down rate.

2. Generation order parameters are firstly introduced to characterize spin-powered X-ray pulsars. Three different GOPs are defined according to the different effects of magnetic and electric field on photon absorption in cascade processes in the context of polar-cap magnetospheric accelerators. The relations of three GOPs and photon index of X-ray pulsars in our sample are displayed. We find the correlation of $\Gamma-\zeta_3$ is better than others, which suggests that in X-ray pulsars the conversion of high energy photons into $e^+/e^-$ through magnetic fields is the dominated mechanism over electric field.

3. We provide a method based on the concept of GOPs in the polar-cap scenario to calculate X-ray luminosity of spin-powered pulsars. In the calculation, we assume an average high energy spectrum, and we also derive it from the observed X-ray and $\gamma$-ray luminosity of 7 $\gamma$-ray pulsars in our sample. The theoretical luminosity versus the observed values is displayed in Fig. 6 with $s=-0.6$. We find that the polar-cap model could well explain most X-ray pulsars except Vela pulsar. The expected relation between the average energy spectrum $s$ and GOP is also presented (though not obviously) by the observed data of 7 $\gamma$-ray pulsars (Fig. 5). 
 
4. From our calculations and analysis, we think that the X-ray emission from three Vela-type pulsars should be out-gap dominated, while Crab-like pulsars and three millisecond pulsars in our sample are polar-cap dominated ones.

5. X-ray luminosity estimated by our mothod versus spin-down power of X-ray pulsars is displayed in Fig. 8. The strong correlation suggests that the primary energy of X-ray emission in spin-powered pulsars comes from the rotational energy loss. Our calculated results are also well consistent with the $L_{\rm X}\sim 10^{-3}L_{\rm sd}$ feature derived by ROAST observations.

In the present paper, we display the correlation between the power-law photon index and spin-down rate for X-ray pulsars which might lies in a deeper physical interpretation. And to check this correlation, we introduce three generation order parameters according to the photon absorption in magnetic and electric fields, and show the relations of $\Gamma-\zeta_{1,2,3}$ which may suggest that the correlation of $\Gamma-|\dot\Omega|$ originates from the magnetosphere acceleration and cascade processes. In fact, the concept of GOPs has been further considered through the different physical mechanisms or in a different scenario. Zhang \& Harding (2000) modified the cascade picture in the polar-cap model by including the inverse Compton scattering (ICS) effect. They took it as the ``full cascade'' scenario, and defined a new form of GOPs to describe the ICS cascade branches. X-ray luminosity based on their GOPs was estimated, and their results also reproduced the observed different dependances of high-energy luminosity on the pulsar spin-down power by ROSAT and ASCA. X-ray emission from millisecond pulsars in their scenario is dominated by thermal components from polar-cap heating, but three millisecond pulsars in our sample with the spectral measurements show the non-thermal components. 

X-ray luminosity of spin-powered pulsars using GOPs in the context of polar-cap model can explain the most X-ray pulsar data well except Vela pulsar. We have found that the outer-gap models (e.g. Wang et al 1998) can match the observed data of the Vela-type pulsars, but the prediction is not well consistent with other pulsars yet. We have suggested that the Vela-type pulsars could be outer-gap dominated objects. Some researchers have argued that the improved outer-gap models could explain all the observed pulsar data (Hirotani \& Shibata 2001, Hirotani et al. 2003). In the same time, the $L_{\rm X}\sim 10^{-3}L_{\rm sd}$ feature could be interpreted by a revised outer-gap model (Cheng, Gil, Zhang 1998). The primary particle energy is determined by the size of outer gap region. They considered a thick outer gap scenario to calculate the non-thermal X-ray luminosity which could meet the observed data well. However, the observed luminosity they took included both thermal and non-thermal components, so Cheng \& Zhang (1999) improved the model by incorporating thermal contributions, specially polar-cap thermal components. Therefore, if we provide that GOPs can describe all pulsars, the GOPs based on outer-gap models could be also defined assuming the primary particle (first generation particle) emissing curvature photons (first generation photons) is accelerated by outer-gap region. These GOPs describing cascade processes in the outer-gap scenario may characterize Vela-type pulsars properly.

X-ray luminosity of spin-powered pulsars concludes thermal and non-thermal components. What component dominated in pulsars and what mechanism responsible for the emission of different classes of pulsars are the interesting and important problems. The GOP concept discussed here can characterize the non-thermal emission of X-ray pulsars. For the resolution limit of the present missions, most pulsars have no spectral measurements, and pulsed and unpulsed luminosities cannot be resolved. So there are only 15 X-ray pulsars in our analysis. More comfirmative conclusion will require high resolution X-ray observations of non-thermal emission and spectral properties of spin-powered pulsars.

\acknowledgments{We are very grateful to the referee for the helpful suggestions that enable us to improve the manuscript significantly, and to Jian Sang and Ye Lu for the help and discussions. We also thank Dr. Z. Zheng and B. Zhang for the fruitful comments. This work is supported by National Natural Science Foundation of China under grant 10273011.}

\begin{figure}
\psfig{figure=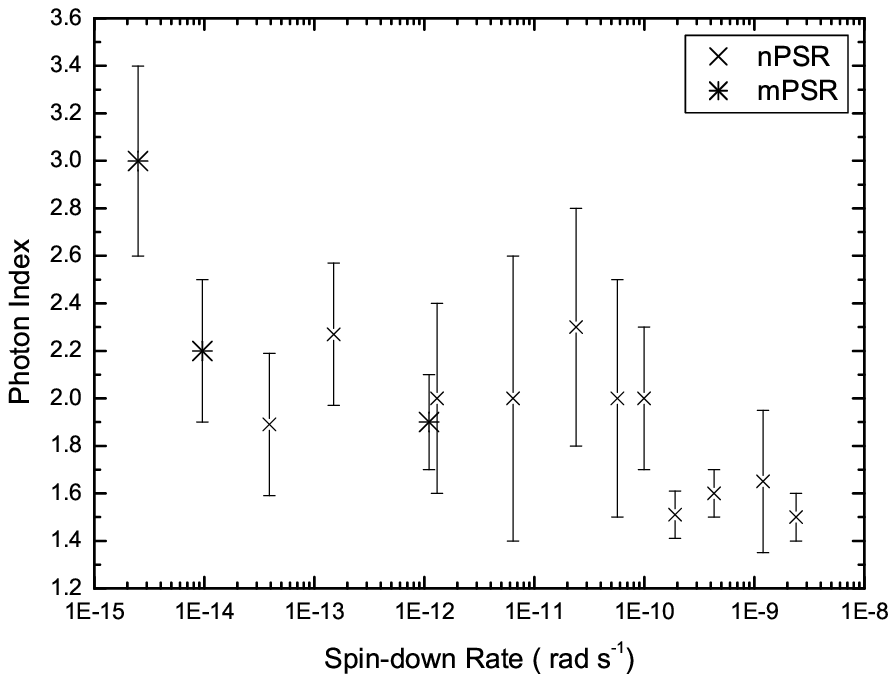,angle=0,width=14cm}
\caption{The variation of power-law photon index versus spin-down rate for spin-powered X-ray pulsars, where nPSR denotes the normal pulsar, mPSR is the millisecond pulsar (the same in the following figures). The photon index decreases with increasing spin-down rate though with a large dipersion.}
\end{figure}

\begin{figure}
\psfig{figure=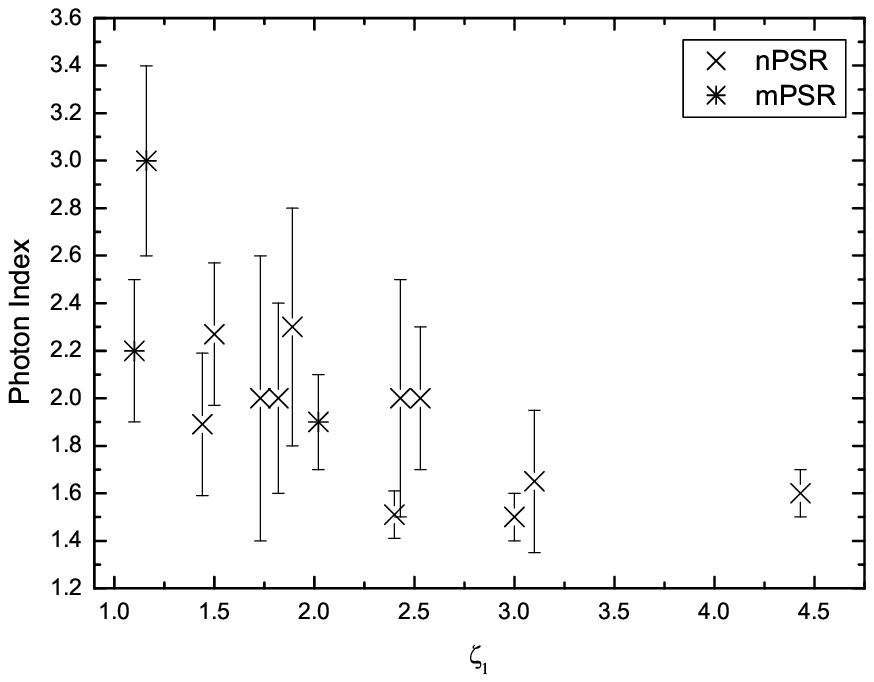,angle=0,width=14cm}
\caption{Photon index $\Gamma$ versus generation order parameter $\zeta_1$.}
\end{figure}

\begin{figure}
\psfig{figure=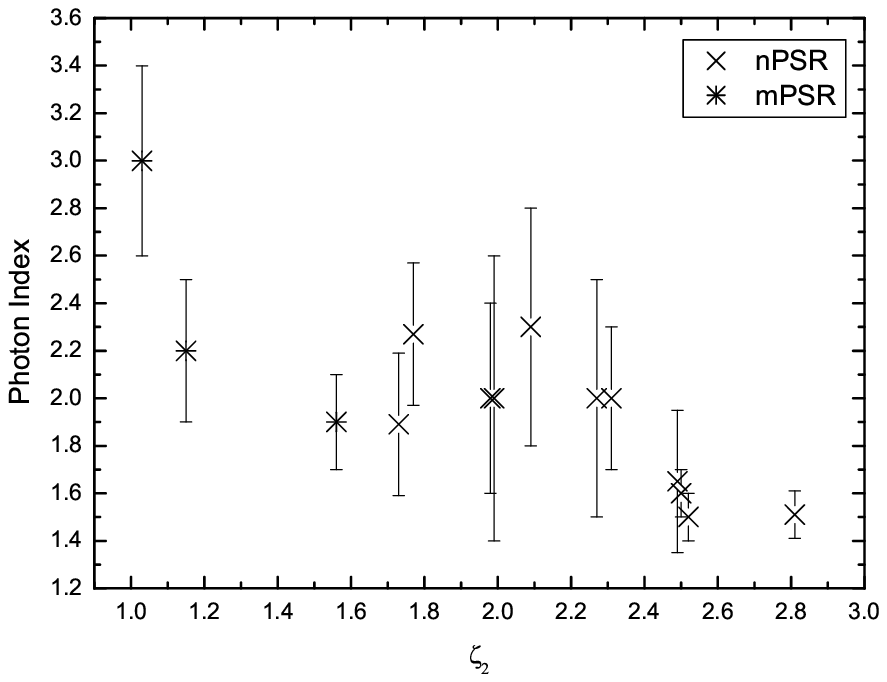,angle=0,width=14cm}
\caption{Photon index $\Gamma$ versus generation order parameter $\zeta_2$.}
\end{figure}

\begin{figure}
\psfig{figure=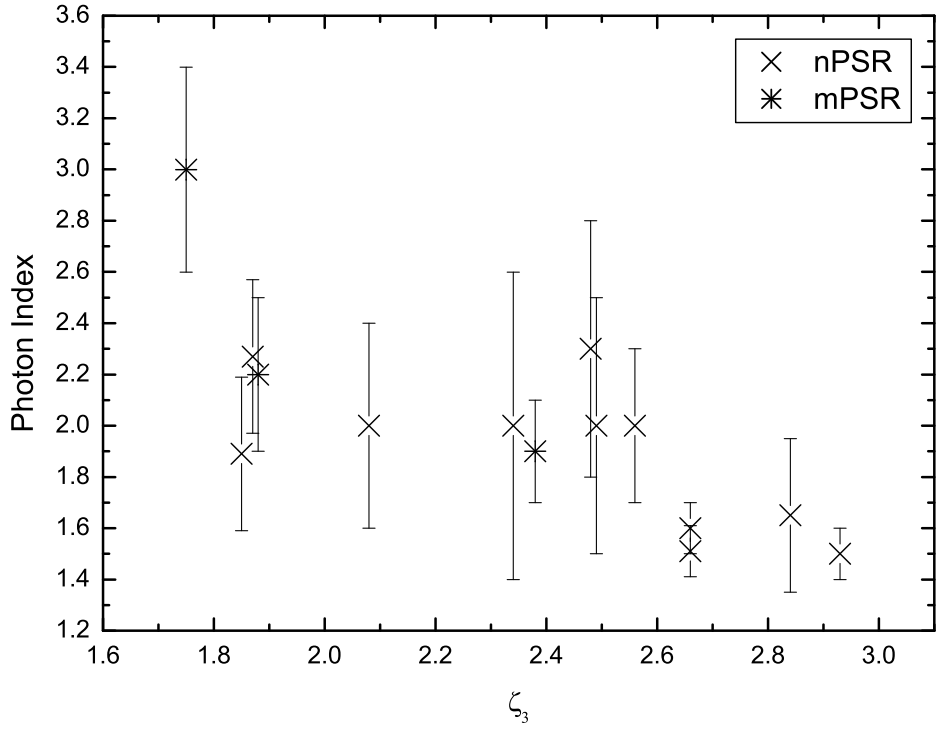,angle=0,width=14cm}
\caption{Photon index $\Gamma$ versus generation order parameter $\zeta_3$. This correlation of $\Gamma-\zeta_3$ is better than these derived from the other two generation order parameters. The diagram shows the harder spectral properties with larger generation order parameters.}
\end{figure}

\begin{figure}
\psfig{figure=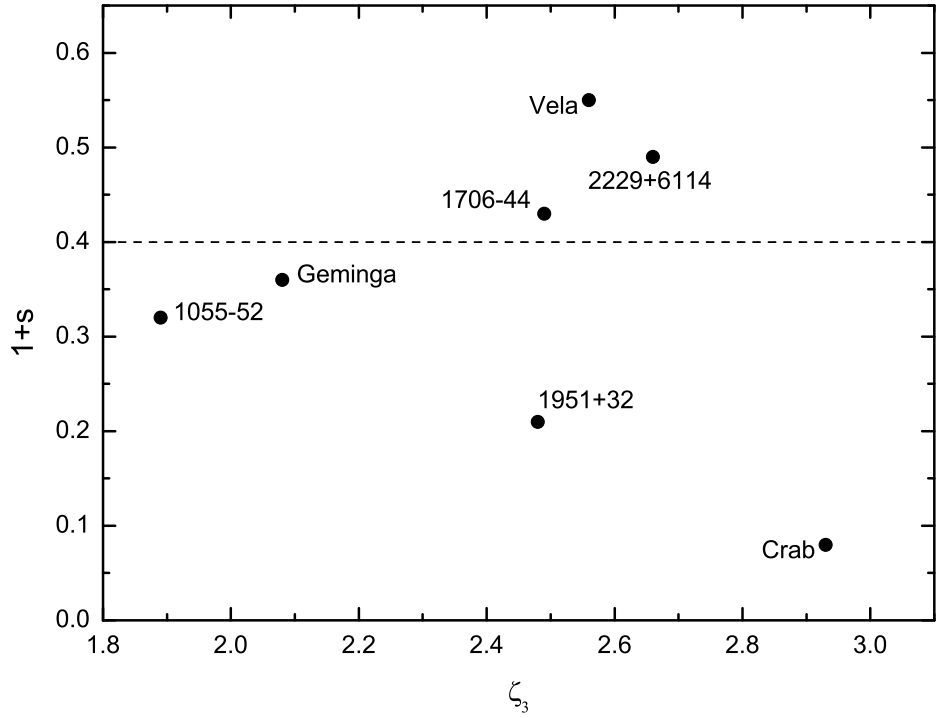,angle=0,width=14cm}
\caption{The average high energy spectrum $(1+s)$ versus the generation order parameter $\zeta_3$ of 6 $\gamma$-ray pulsars. The dashed line denotes $s=-0.6$. The data points are really around the line, comfirming the consistancy of our luminosity estimation. If comparing Crab with other pulsars, a weak relation between $s$ and $\zeta_3$ exists, softer spectra with larger generation order parameters.}
\end{figure}

\begin{figure}
\psfig{figure=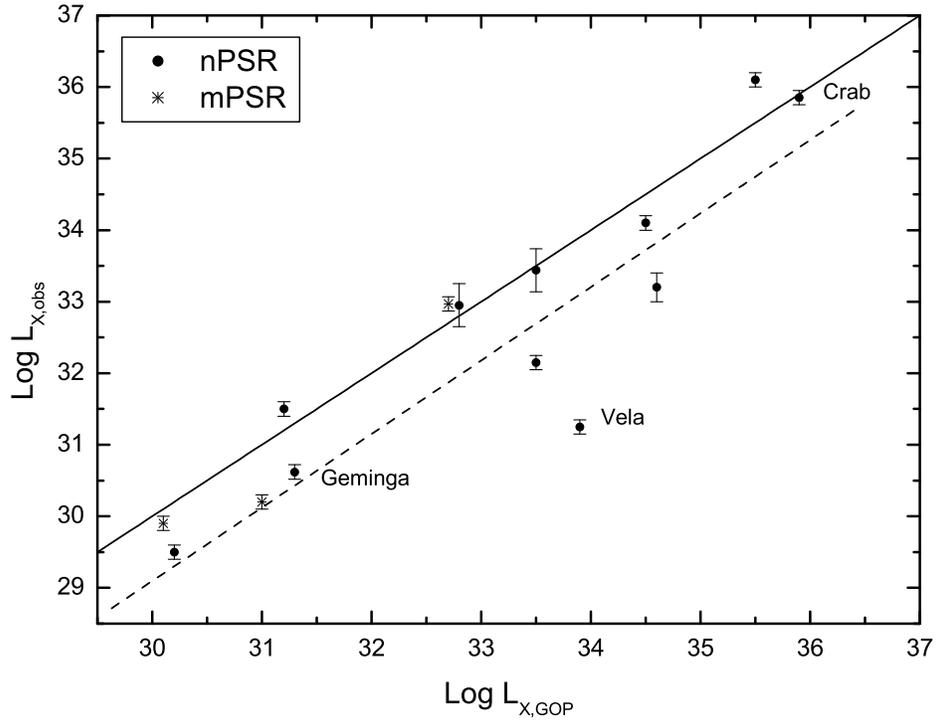,angle=0,width=14cm}
\caption{The observed X-ray luminosity versus the predicted one by our polar-cap model, and $s=-0.6$. The solid line represents the diagonal line $L_{\rm X,obs}=L_{\rm X,GOP}$, the dashed line is $L_{\rm X,obs}=0.02L_{\rm X,GOP}^{1.02}$.}
\end{figure}

\begin{figure}
\psfig{figure=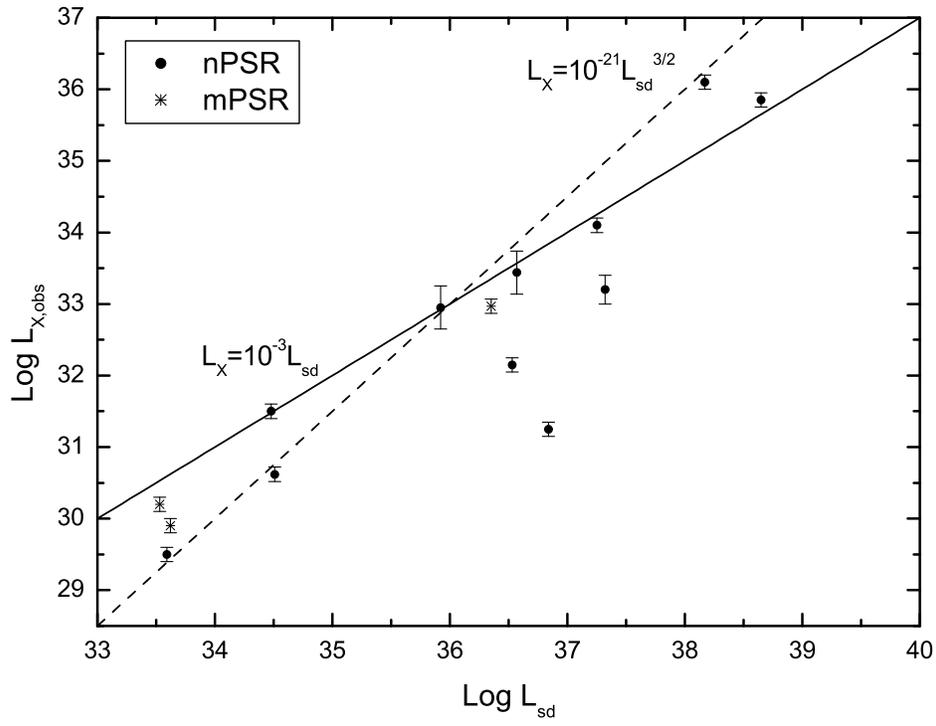,angle=0,width=14cm}
\caption{The observed X-ray luminosity versus total spin-down energy for 15 spin-powered pulsars. The solid line represents $L_{\rm X}=10^{-3}L_{sd}$, and the dashed line is $L_{\rm X}=10^{-21}L_{sd}^{3/2}$.}
\end{figure}

\begin{figure}
\psfig{figure=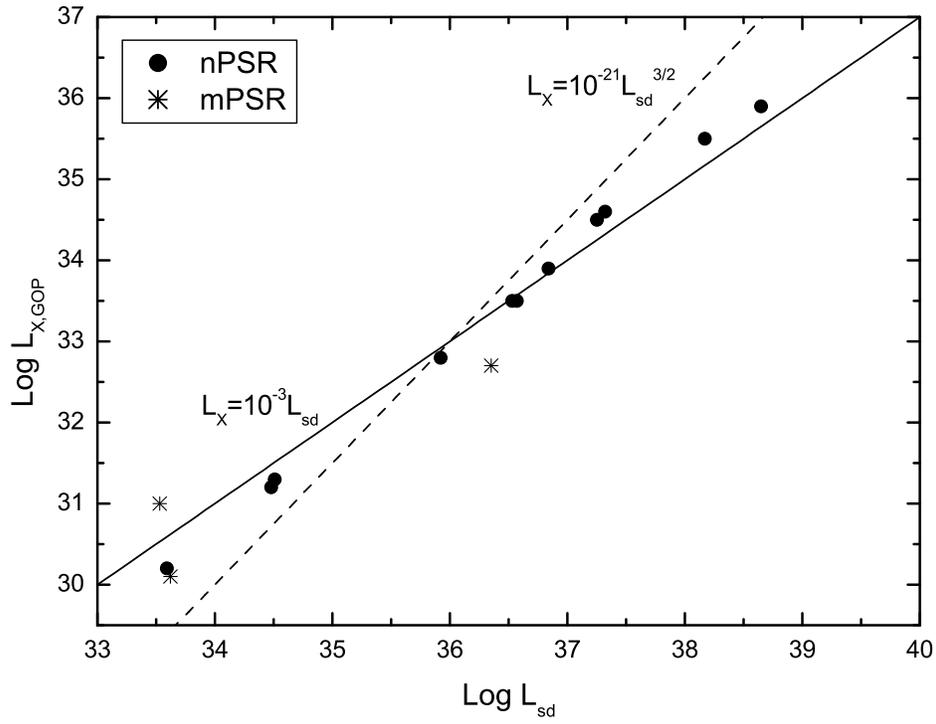,angle=0,width=14cm}
\caption{X-ray luminosity estimated by polar-cap models with $s=-0.6$ versus total spin-down energy for 15 spin-powered pulsars. The solid line represents $L_{\rm X}=10^{-3}L_{sd}$, and the dashed line is $L_{\rm X}=10^{-21}L_{sd}^{3/2}$.}
\end{figure}

\begin{table}
\caption{The characteristics of some spin-powered X-ray pulsars}
\begin{center}
\begin{tabular}{l c c c c c c c c c c l}
\tableline
\tableline
PSR & $\Gamma$ & $P$ & $\dot P$ & $\zeta_1$ & $\zeta_2$ & $\zeta_3$ & log$L_{\rm X}^b$ & log$L_{\rm X,GOP}$ & log$L_{sd}$ & Ref.$^c$ \\
\tableline
Crab  &  $1.5\pm 0.1$ &  0.033 &  4.201e-13 &  3.00 &  2.52 &  2.93  &  $35.85\pm 0.1$ &  35.9 & 38.65 & 1,2 \\
1509-58  &  $1.6\pm 0.1$  &  0.150 &  1.54e-12 &  4.43 &  2.50 &  2.66 & $34.10\pm 0.1$ &  34.5 & 37.25 & 2,3 \\
0540-69 &  $1.65\pm 0.3$ & 0.050 &  4.79e-13 & 3.10 &  2.49 &  2.84  &  $36.1\pm 0.1$  &  35.5 &  38.17 & 2,4 \\
Vela & $2.0\pm 0.3$ &  0.089 &  1.247e-13 & 2.53 &  2.31 &  2.56 & $31.25\pm 0.1$ &  33.9 &  36.84 & 2,5,10 \\
1706-44  &  $2.0\pm 0.5$ &  0.102 &  9.3e-14 &   2.43 &  2.27 &  2.49  &  $32.15\pm 0.1$ & 33.5 &  36.53 & 2,6,7 \\
1951+32 &  $2.3\pm 0.5$ &  0.0395 & 5.85e-15  & 1.89  & 2.09  & 2.48  &  $33.44\pm 0.3$ &  33.5 & 36.57 & 2,10 \\
1259-63 & $2.0\pm 0.6$ &  0.0478 & 2.27e-15  & 1.73  & 1.99  & 2.34  &  $32.95\pm 0.3$  & 32.8 & 35.92 & 2,10 \\
Geminga &  $2.0\pm 0.4$ &  0.237 &  1.1e-14  &  1.82 &  1.98 &  2.08 & $30.62\pm 0.1$ &  31.3 & 34.51 & 2,8 \\
1929+10 &  $2.27\pm 0.3$  & 0.227 &  1.16e-15 &  1.50 &  1.77 &  1.87 & $29.5\pm 0.1$  & 30.2 &  33.59 & 2,10 \\
2227+6122 &  $1.51\pm 0.1$ & 0.0516 & 7.83e-14 &  2.40  & 2.81  & 2.66  &  $33.2\pm 0.2$ &  34.6 & 37.32 & 9,11 \\
1055-52 & $1.89\pm 0.3$ &  0.53 & 1.72e-15 &  1.44 &  1.73 & 1.85 &  $31.5\pm 0.1$ & 31.2 &  34.48 & 10 \\
1821-24$^a$ &  $1.9\pm 0.2$ &  0.003  & 1.62e-18 &  2.02  & 1.56 &  2.38  &  $32.97\pm 0.1$ & 32.7 &  36.35 & 10,12 \\
0437-47$^a$ &  $2.2\pm 0.3$ & 0.005 &  3.8e-20  &  2.45 &  1.15 &  1.88  &  $29.9\pm 0.1$ &  30.1 &  33.62 & 10,13 \\
0030+0451$^a$ &  $3.0\pm 0.4$ & 0.005 &  1.0e-20 & 1.16 &  1.03 &  1.75 &  $30.2\pm 0.1$ &  31.0 &  34.48 & 10 \\
\tableline
\end{tabular}
\end{center} 
\tablecomments{$\Gamma$ is the power-law photon index, $P$ is the period in units of s and $\dot P$ in units of s s$^{-1}$. $\zeta_{1,2,3}$ is the generation order parameters with three different definitions. The luminosity is in units of erg\ s$^{-1}$. $L_{\rm X}$ is the observational X-ray luminosity $L_{\rm X,GOP}$ is the luminosity given by our polar-cap model with $s=-0.6$, and $L_{sd}$ is the pulsar's total spin-down power. \\
$^a$ The pulsars are millisecond pulsars (mPSR), others are normal pulsars (nPSR). \\
$^b$ All luminosities $L_x$ are calculated for the ROSAT energy range 0.1-2.4 keV in isotropy. \\
$^c$ References: 1. Toor \& Seward (1977); 2. Becker \& Tr\"umper (1997); 3. Kawai et al. (1992); 4. Finley et al. (1993); 5. Harding et al. (2002); 6. Becker et al. (1995); 7. Gotthelf et al. (2002); 8. Halpern \& Ruderman (1993); 9. Halpern et al. (2001a); 10. Becker \& Aschenbach (2002); 11. Halpern et al. (2001b); 12. Saito et al. (1997); 13. Zavlin et al. (2002).
}
\end{table}

\end{document}